\documentclass[aps, prx, reprint, superscriptaddress, longbibliography, floatfix]{revtex4-2}
\usepackage{graphicx}
\usepackage{dcolumn}
\usepackage{bm}
\usepackage{physics}
\DeclareMathAlphabet{\pazocal}{OMS}{zplm}{m}{n}
\usepackage{amsmath}
\usepackage{amsfonts}
\usepackage{amssymb}
\usepackage{mathrsfs}
\usepackage{subfigure}
\usepackage{color}
\usepackage{xcolor}
\usepackage{tikz}
\usepackage[colorlinks=true,linkcolor=blue,anchorcolor=red,citecolor=blue, urlcolor=blue]{hyperref}

\newcommand{\occsite}{\tikz[baseline=-0.5ex]\fill (0,0) circle (2pt);}
\newcommand{\empsite}{\tikz[baseline=-0.5ex]\draw (0,0) circle (2pt);}
\newcommand{\occsiteRed}{\tikz[baseline=-0.5ex]\fill[red] (0,0) circle (2pt);}
\newcommand{\empsiteRed}{\tikz[baseline=-0.5ex]\draw[red] (0,0) circle (2pt);}

\begin{document}
\preprint{APS/123-QED}
 
\title{Anomalous coarsening and nonlinear diffusion of kinks in an one-dimensional quasi-classical Holstein model}

\author {Ho Jang}
\affiliation{Department of Physics, University of Virginia, Charlottesville, Virginia, 22904, USA}

\author {Yang Yang}
\affiliation{Department of Physics, University of Virginia, Charlottesville, Virginia, 22904, USA}

\author {Gia-Wei Chern}
\affiliation{Department of Physics, University of Virginia, Charlottesville, Virginia, 22904, USA}

\begin{abstract}
We study the phase-ordering dynamics of a quasi-classical Holstein model. At half-filling, the zero-temperature ground state is a commensurate charge-density-wave (CDW) with alternating occupied and empty sites. This quasi-classical formulation enables us to isolate the role of electrons in coarsening dynamics. Following a thermal quench, CDW domains grow through the diffusion and annihilation of kinks -- topological defects separating the two symmetry-related CDW orders. While standard diffusive dynamics predicts domain sizes scaling as the square root of time, our large-scale simulations reveal a slower power-law growth with a temperature-dependent exponent. We trace this anomalous behavior to a cooperative kink hopping arising from Fermi-Dirac statistics of electrons and quasi-conservation of electron numbers. The correlated-hopping of kinks in turn gives rise to an effective diffusion coefficient that depends on the kink density. These results uncover a new mechanism for slow coarsening and carry implications for phase-ordering in the full Holstein model and related electron-phonon systems.
\end{abstract}


\date{\today}
\maketitle

\section{Introduction}

\label{sec:intro}

Coarsening refers to a process in which small structures or domains within a system merge or grow over time, leading to the formation of larger, more stable structures~\cite{Cugliandolo15}. This phenomenon is observed in a wide variety of systems, from materials science to cosmology. The study of coarsening dynamics is closely tied to phase ordering dynamics, an important subject in nonequilibrium physics with a rich historical foundation~\cite{Bray1994,Onuki2002,Puri2009}.  Several empirical dynamics for the order parameter fields, based on symmetry considerations and conservation laws, have been developed~\cite{hohenberg77}. The systematic classification of empirical order-parameter dynamics forms the foundation for categorizing different universality classes of phase-ordering dynamics~\cite{Bray1994,Onuki2002}.

The different universality classes of coarsening dynamics are best illustrated by the time-dependent correlation length, $\xi(t)$, of the order-parameter field. This length scale $\xi$ also provides a measure of the characteristic size of the ordered domains. In the late stage of coarsening, the correlation length increases following a power-law dependence on time
\begin{eqnarray}
	\xi \sim t^{1/z},
\end{eqnarray}
where the growth exponent is related to the dynamical exponent $z$. For systems characterized by a broken $Z_2$ or Ising symmetry in $d \ge 2$ dimensions, there are two dynamical universality classes: one with $z = 2$ for non-conserved and conserved Ising order parameters, and the other with $z=3$ for conserved Ising orders. The square-root of time scaling of correlation length, also known as the Allen-Chan growth law for non-conserved Ising order coarsening, is linked to a curvature-driven mechanism of domain expansions, underscoring the crucial role of topological defects in coarsening dynamics. Indeed, the dynamical evolution of topological defects of the order-parameter field provides a unifying framework for understanding the different universality classes in phase-ordering dynamics~\cite{Bray1994}.

For non-conserved Ising order in one dimension (1D), the coarsening dynamics also follows a power law with a dynamical exponent $z=2$, albeit driven by a different mechanism. Unlike the curvature-driven domain-wall motion in higher dimensions, the coarsening of Ising domains in 1D occurs through the diffusion and annihilation of kinks, which are particle-like topological defects characteristic of 1D Ising systems. The topological nature of kinks arises from their role as junctions between ordered domains with opposite Ising order. As a result, the annihilation of two diffusive kinks leads to the disappearance of the Ising domains they bound. The diffusive motion of kinks,  modeled by unbiased random walks, results in a linear scaling of the squared displacement $\langle \Delta x^2 \rangle \sim t$. As kinks within the a range $\xi \sim \sqrt{ \langle \Delta x^2 \rangle}$ tend to annihilate with each other, this scaling thus gives rise to a dynamical exponent of $z = 2$~\cite{Toussaint83,Torney83,Krapivsky_book,Ovchinnikov1989, Hughes1995RandomWalks}. This diffusion-controlled annihilation scenario of 1D domain coarsening has been confirmed by exact solutions for both Ising systems and generalizations such as Potts and voter models~\cite{Cox89,Amar90,Sire95,Derrida95,Derrida96}. Beyond their role in coarsening, kinks also underpin a broad range of other nonequilibrium dynamical phenomena in one-dimensional systems~\cite{Mayo2021, krapivsky_slow_2010,kibble1976,zurek1996,Jeong2020}.

For electron systems on bipartite lattices, charge density waves (CDWs) with staggered electron densities on the two sublattices represent one of the simplest realizations of an Ising order~\cite{Gruner1988,Gruner1994,kivelson03,fradkin15,tranquada95}. The unequal electron densities in the doubled unit cell of CDW states leads to a broken $Z_2$ symmetry. CDWs are one of the most common symmetry-breaking phases in functional electron materials, often coexisting or competing with other electronic orders such as spin density waves or superconductivity. Despite their prevalence in many materials systems, coarsening dynamics of CDWs has yet to be carefully examined. In particular, the nontrivial interplay between the slow CDW order parameter and the fast quasi-particles could lead to unexpected dynamical behaviors. Indeed, recent works have uncovered anomalous CDW coarsening in 2D that cannot be described by the conventional Allen-Cahn growth law~\cite{Cheng23}. 

\begin{figure}
\includegraphics[width=0.99\columnwidth]{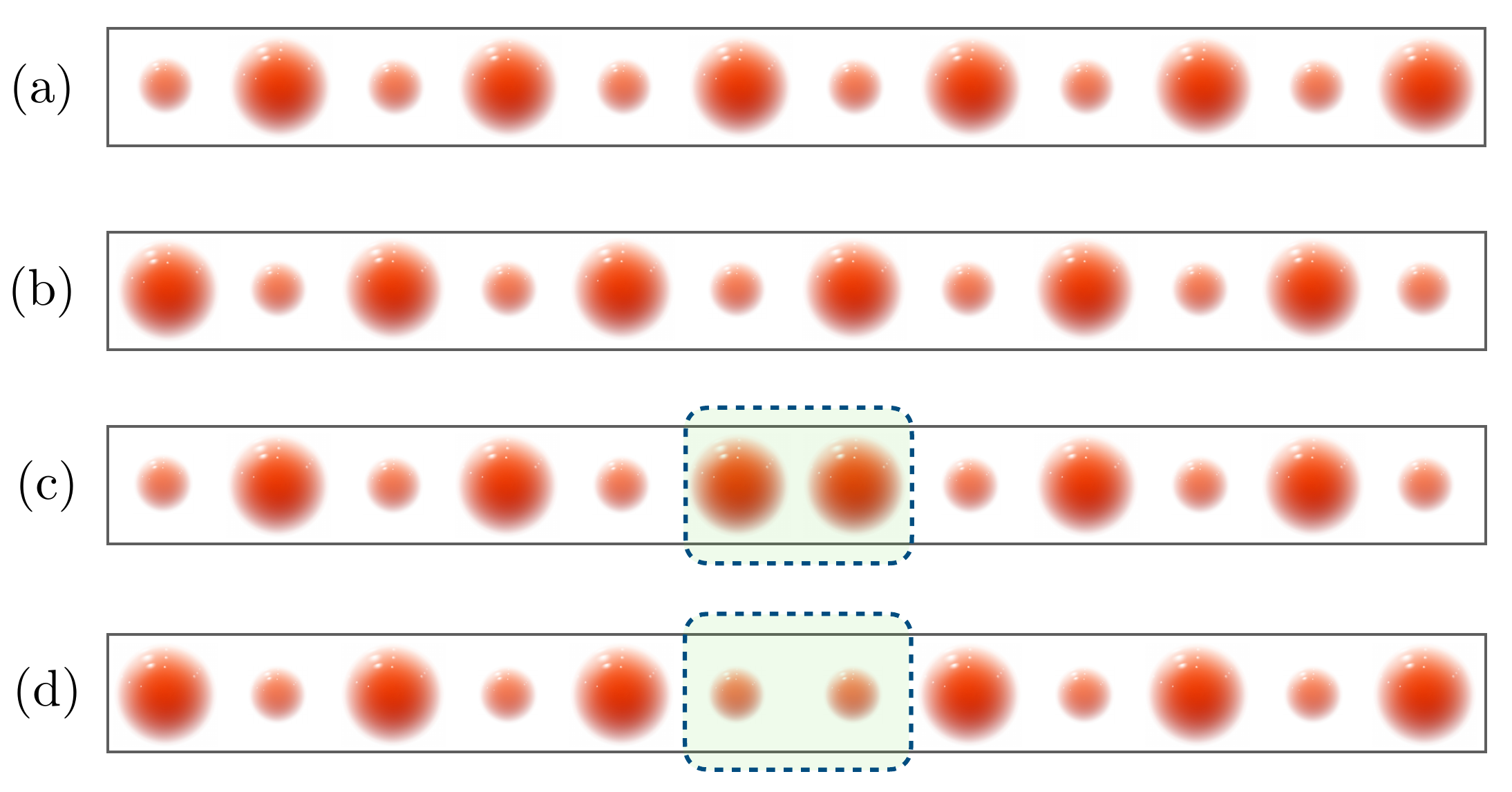}
\caption{CDW order in 1D and kinks. The electron density modulation is described by a scalar order parameter $\Delta$ as $n_i = \overline{n} + \Delta (-1)^i$. The two degenerate ground states in panels~(a) and (b) correspond to opposite signs of $\Delta$. In an inhomogeneous CDW states, domains of opposite CDW order are connected by topological defects, known as kinks, as illustrated in panels~(c) and (d). }
\label{fig:cdw-order}  
\end{figure}

For 1D systems, the electron density modulation in a CDW is characterized by a scalar order parameter $\Delta$ as: $n_i = \overline{n} + \Delta (-1)^{i}$, where $\overline{n}$ is the average electron density. This leads to a doubled unit cell with the two inequivalent sites having different electron numbers $n = \overline{n} \pm \Delta$; see Fig.~\ref{fig:cdw-order}(a) and (b). The two symmetry-related CDW ground states correspond to positive and negative values of~$\Delta$. In an inhomogeneous CDW state, domains of opposite CDW order are connected by kinks, as shown in Fig.~\ref{fig:cdw-order}(c) and (d), which are topological defects that must be created and annihilated in pairs. As discussed above, the coarsening of CDW domains in 1D is determined by the diffusive motions of these topological defects. As the CDW order is stabilized by the electron-lattice coupling, the non-local nature of the electronic subsystems could impose unusual constraints  on the diffusive motions of kinks and coarsening behaviors.

In this work, we present a comprehensive study of CDW coarsening in a one-dimensional system stabilized by Holstein-type electron-phonon coupling. We begin by investigating the ordering dynamics of the CDW phase in the standard 1D Holstein model with classical lattice degrees of freedom. At half-filling, the ground state of this semi-classical system exhibits CDW order with a doubled unit cell, as illustrated in Fig.~\ref{fig:cdw-order}. In the adiabatic Born-Oppenheimer regime, the system's dynamics are described by a Langevin equation for the lattice distortion, with driving forces arising from the electronic subsystem. Our large-scale dynamical simulations---where the electronic states are computed via exact diagonalization at each time step---reveal an unconventional power-law coarsening of CDW domains, which deviates significantly from the conventional diffusive growth law.

To investigate the origin of the anomalous coarsening, we analyze a simplified yet nontrivial {\em quasi-classical} limit of the Holstein model obtained by setting the electron hopping to zero. In this limit, the electron subsystem can be solved exactly, since its eigenenergies are determined directly by the on-site lattice distortions. Despite this apparent simplicity, the system exhibits relaxation dynamics strikingly similar to those of the full Holstein model: domain growth follows a power law with a temperature-dependent exponent that substantially exceeds the diffusive limit $z=2$. We trace this anomalous scaling to cooperative kink hopping constrained by electron-number conservation, which makes the effective diffusion coefficient explicitly dependent on kink density. This mechanism uncovers a new pathway to anomalously slow coarsening and suggests broader implications for phase-ordering dynamics in the Holstein model and other electron-phonon coupled systems.

\section{Coarsening of CDW order in Holstein model}

\label{sec:semiclassical-Holstein}

We begin by considering the 1D semi-classical Holstein Hamiltonian~\cite{Holstein59,holstein1959b,scalettar1989} for spinless fermions and classical phonons at half-filling:
\begin{eqnarray}
    & & \hat{\mathcal{H}} = -t_{\rm nn}\sum_{i}(\hat{c}_i^\dagger \hat{c}_{i+1} + {\rm h.c.}) -g\sum_i \left(\hat{c}_i^\dagger \hat{c}_i - \frac{1}{2}\right)Q_i \nonumber \\
    & & \qquad +\sum_i \left(\frac{P_i^2}{2m}+\frac{kQ_i^2}{2}\right)+\kappa \sum_{i} Q_iQ_{i+1}.   
\label{eqn:Holstein1D}
\end{eqnarray}
Here, $\hat{c}_i^\dagger$ ($\hat{c}_i$) denote the creation (annihilation) operators of an electron at site $i$, with hopping amplitude $t_{\rm nn}$ between nearest-neighbor sites. ${Q}_i$ represents the local phonon displacement, and ${P}_i$ is its conjugate momentum. The first term corresponds to electron hopping between adjacent lattice sites with amplitude $t$. The second term describes the onsite electron–phonon interaction, involving the electron number operator $\hat{n}_i = \hat{c}_i^\dagger \hat{c}_i$ coupled to the local phonon displacement ${Q}_i$ with strength $g$. The third term represents the Einstein phonon contribution, modeled as independent harmonic oscillators at each lattice site, characterized by mass $m$ and spring constant $k$. The final term accounts for nearest-neighbor phonon–phonon coupling, which stabilizes vibrational modes with coupling strength $\kappa$.

The Holstein model has long served as a fundamental framework for investigating CDW phase transitions, as well as a range of phenomena arising from electron–phonon interactions, including polaron dynamics and superconductivity~\cite{noack91,zhang19,chen19,hohenadler19}. At half-filling on bipartite lattices, such as the square and honeycomb, the model exhibits a finite-temperature transition into a CDW-ordered phase. Symmetry considerations indicate that this transition should belong to the Ising universality class, a prediction that has been confirmed by quantum Monte Carlo (QMC) simulations~\cite{noack91,zhang19}. The CDW phase of the half-filled Holstein model can therefore be characterized by an Ising-type order parameter field~\cite{mcmillan75}.

From a dynamical perspective, however, recent large-scale simulations enabled by machine-learning methods~\cite{Cheng23} have revealed that the coarsening of CDW domains in two dimensions deviates from the Allen–Cahn growth law expected for non-conserved Ising-type order parameter fields. This anomalous coarsening behavior has been tentatively attributed to the unusual geometry of domain walls~\cite{Cheng23}, which is incompatible with the curvature-driven mechanism underlying Allen–Cahn dynamics. A complete theoretical understanding of the coarsening dynamics of the Ising-type CDW order nevertheless remains an open challenge.

The CDW order in a 1D Holstein chain, which is also a bipartite lattice, is schematically depicted in FIG.~\ref{fig:cdw-order}(a) and (b). In one dimension, long-range Ising order is destroyed at finite temperatures by the proliferation of kinks. Consequently, the coarsening dynamics in such systems pertains to the growth of Ising domains prior to saturation of the correlation length, whose equilibrium value is set by the inverse of the kink density. As discussed in Sec.~\ref{sec:intro}, standard 1D Ising systems exhibit power-law domain growth with a dynamic exponent $z = 2$, superficially resembling the Allen–Cahn behavior observed in dimensions $D \geq 2$. The mechanism driving this growth, however, is entirely different: rather than curvature-driven relaxation, it originates from the reaction-diffusion dynamics of kinks.

Here we investigate the coarsening dynamics of charge-density waves (CDWs) in the one-dimensional half-filled Holstein model, focusing on its adiabatic dynamical limit. Within the semiclassical approximation, the dynamics of the local lattice modes are governed by an effective Langevin equation~\cite{ermak80,glauber_time-dependent_1963,hanggi1982}:
\begin{eqnarray}
	\label{eq:langevin}
	m\frac{d^2Q_i}{dt^2} = - \frac{\partial \langle  \mathcal{H} \rangle}{\partial Q_i}  - \gamma \frac{dQ_i}{dt} + \eta_i(t),
\end{eqnarray}
where a Langevin thermostat is employed to incorporate the influence of a thermal reservoir during phase ordering, where $\gamma$ denotes the damping constant and $\eta_i(t)$ represents a zero-mean thermal noise characterized by its correlation functions.
\begin{eqnarray}
	 \langle \eta_i(t) \rangle &=& 0, \\
	 \langle \eta_i(t) \eta_j(t') \rangle &=& 2 \gamma k_B T \delta_{ij} \delta(t - t'). \nonumber
\end{eqnarray}
A key step in integrating the Langevin equation is the calculation of the force which can be naturally decomposed into two components: 
\begin{eqnarray}
	\label{eq:force-total}
	& & F_i \equiv - \frac{\partial \langle \mathcal{H} \rangle}{\partial Q_i} = F^{\rm elastic}_i + F^{\rm elec}_i.
\end{eqnarray}
The first term corresponds to the classical elastic restoring force arising from lattice distortions:
\begin{eqnarray}
	\label{eq:F_elastic}
	F^{\rm elastic}_i = - k Q_i - \kappa \sum_{j \in \mathcal{N}(i)}  Q_j,
\end{eqnarray}
where $\mathcal{N}(i)$ denotes the two nearest neighbors of site-$i$. The second electronic contribution can be evaluated using the Hellmann-Feynman theorem~\cite{hellmann1937,feynman1939}, $\partial \langle \mathcal{H}_e \rangle / \partial Q_i = \langle \partial \mathcal{H}_e / \partial Q_i \rangle$, which gives
\begin{eqnarray}
	\label{eq:F_elec}
	F^{\rm elec}_i = g \left(  \langle \hat{n}_i \rangle  - \frac{1}{2} \right),
\end{eqnarray}
where $\hat{n}_i  =  \hat{c}^\dagger_i \hat{c}^{\,}_i $ is the electron number operator and $\langle \cdots \rangle$ denotes its quasi-equilibrium expectation value.  The electron force is proportional to the deviation of the average local electron density from half-filling.

\begin{figure}
\includegraphics[width=0.99\columnwidth]{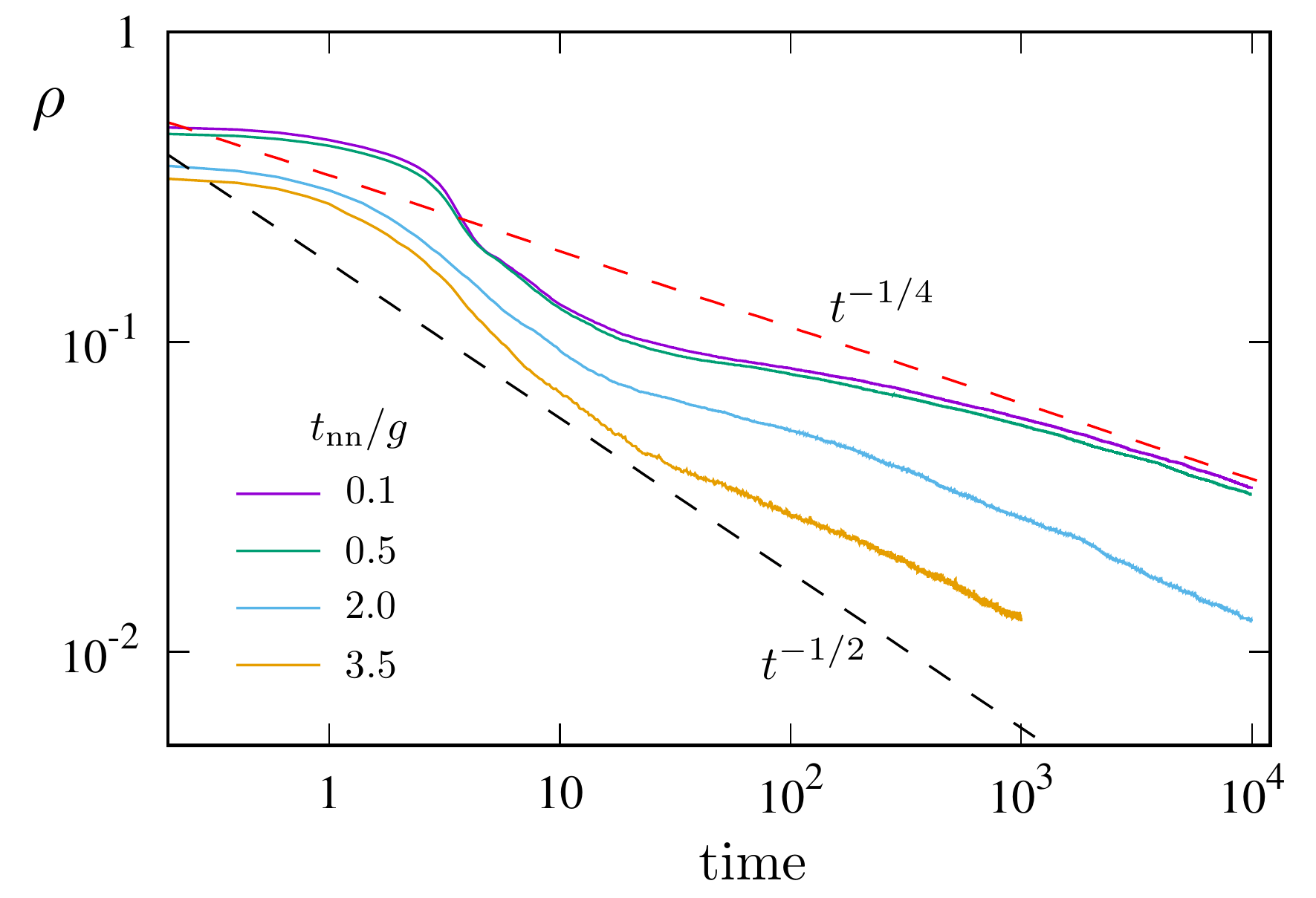}
\caption{Density of kinks $\rho$ versus time of the 1D semi-classical Holstein model with various hopping amplitudes $t_{\rm nn}$ at the temperature $T = 0.032$.}
\label{fig:nk-t-quantum}  \
\end{figure}

The growth of domains during phase ordering is typically much slower than the relaxation of the electronic subsystem. As a result, the evolution of the CDW state can be treated within the adiabatic approximation, which assumes a clear separation between the slow lattice dynamics and the fast electron relaxation. This approximation is formally equivalent to the Born-Oppenheimer scheme widely used in quantum molecular dynamics simulations~\cite{marx2009}. Within this framework, the electronic subsystem is assumed to remain in quasi-equilibrium with the instantaneous lattice configuration, allowing expectation values--such as the local electron density--to be evaluated from a Boltzmann distribution:
\begin{eqnarray}
	 \langle \hat{n}_i \rangle  = \frac{1}{Z_e} {\rm Tr}\left( \hat{c}^\dagger_i \hat{c}^{\,}_i  \, e^{-\beta \mathcal{H}_e(\{Q_i\})} \right),
\end{eqnarray}
where $Z_e = {\rm Tr}e^{-\beta \mathcal{H}_e(\{Q_i\})}$ is the partition function of quasi-equilibrium electrons. Because the electronic Hamiltonian is quadratic in the fermionic creation and annihilation operators, the resulting tight-binding model can be solved exactly in real space using the exact diagonalization (ED) method. However, since ED must be performed at every time step of the Langevin dynamics, large-scale coarsening simulations of the Holstein model become computationally demanding. In practice, for one-dimensional systems, we are able to simulate chains of up to $N = 500$ sites.

We performed Langevin dynamics simulations to investigate the relaxation of a one-dimensional Holstein chain following a thermal quench. Specifically, the system was initialized with a random configuration of $Q_i$ and then suddenly coupled to a thermal reservoir at a low temperature $T = 0.032$, with energies measured in units of the nearest-neighbor hopping $t_{\rm nn}$. After a brief transient stage, during which local charge modulation develops, the system evolves into a configuration characterized by well-formed CDW domains separated by domain walls (kinks). To monitor the coarsening of these CDW domains, we compute from simulations the kink density $\rho(t)$ as a function of time, which provides a direct measure of the correlation length $\xi \sim 1/\rho$. The results are shown in Fig.~\ref{fig:nk-t-quantum} for varying ratio of electron-phonon coupling and electron hopping amplitude. Unit can be found in the Appendix~\ref{apd:DimensionalAnalysis}. The annihilation of kinks at late stage exhibits a power-law behavior 
\begin{eqnarray}
	\rho \sim 1/t^{1/z}.
\end{eqnarray}
However, when compared with the $z = 2$ scaling expected from conventional diffusion-limited annihilation (black dashed line in Fig.~\ref{fig:nk-t-quantum}), significant deviations are observed across all coupling strengths~$g$. For instance, the late-stage coarsening at small~$g$ is more accurately characterized by an exponent of $z \approx 4$. While a detailed and systematic investigation is deferred to future work, our present goal is to gain insight into the mechanism underlying the observed anomalous coarsening. To this end, we introduce a further simplified model that permits more tractable analytical analysis of the coarsening dynamics.

\section{Coarsening of CDW order in Qausi-Classical Holstein model}

\label{sec:quasi-classical-Holstein}

Our study of coarsening dynamics in the 1D Holstein model reveals increasingly pronounced deviations from diffusive behavior ($z = 2$) as the electron-phonon coupling strength grows. This trend points to markedly anomalous coarsening in the strong-coupling limit, $g / t_{\rm nn} \to \infty$ or equivalently $t_{\rm nn} \to 0$. To this end, we consider a modified Holstein Hamiltonian without electron hopping:
\begin{eqnarray}
    & & \hat{\mathcal{H}}\,=\,-g\sum_i \left(\hat{n}_i - \frac{1}{2}\right)Q_i  \\
    & & \qquad +\sum_i \left(\frac{P_i^2}{2m}+\frac{kQ_i^2}{2}\right)+\kappa \sum_{\langle ij \rangle} Q_iQ_j. \nonumber
\label{eq:quasi-classical-H}
\end{eqnarray}
The electron part of the Hamiltonian is effectively classical, as it can be trivially diagonalized:
\begin{eqnarray}
	\label{eq:H_elec}
	\hat{\mathcal{H}}_e = -g \sum_i Q_i \hat{n}_i
\end{eqnarray}
This follows from the fact that it commutes with all on-site electron number operators, $[\hat{\mathcal{H}}, \hat{n}_i ] = 0$, and that the number operators themselves commute, $[\hat{n}_i, \hat{n}_j ] = 0$. As a result, eigenstates of $\hat{\mathcal{H}}$ are simultaneous eigenstates of the full set of $\hat{n}_i$, and can be written as Fock states $|n_1, n_2, n_3, \cdots \rangle$ where the eigenvalues are $n_i = 0$ or 1. 

Despite the apparent simplicity of this classical-like Hamiltonian, a nontrivial quantum feature--the Pauli exclusion principle--is retained in its eigenstates. In the absence of a more precise term, we refer to this as a {\em quasi-classical} Holstein model. At finite temperature, the electronic state is determined by the Fermi-Dirac distribution over the single-electron spectrum, which depends directly on the lattice configuration via $\epsilon_i = - g Q_i$. Specifically, the on-site electron number is given by
\begin{eqnarray} \label{eq:quasiclassical-electrondensity}
	\langle \hat{n}_i \rangle = \frac{1}{\exp{-(g Q_i + \varepsilon_F)/T} + 1},
\end{eqnarray}
where the Fermi level $\varepsilon_F$ is determined from the half-filling condition $\sum_i \langle \hat{n}_i \rangle = N/2$. 

On the other hand, electron hopping is essential for the emergence of charge density wave (CDW) states, as exemplified by the Peierls instability, which arises due to Fermi surface nesting and the associated electronic energy gain upon lattice distortion. This mechanism becomes particularly relevant in the weak electron-phonon coupling regime ($g \ll t_{\rm nn}$), where the kinetic energy of electrons dominates. In the strong coupling regime, the commensurate CDW order with a wave vector $q = \pi$ can be understood as arising from a dominant antiferro-distortive interaction mediated by electrons. Here, in order to stabilize the commensurate $q = \pi$ CDW order in the $t_{\rm nn} \to 0$ limit of the quasi-classical Holstein model, an nearest-neighbor antiferro-distortive interaction  $\kappa > 0$, the last term in Eq.~(\ref{eq:quasi-classical-H}), is explicitly included in the Hamiltonian.

\begin{figure*}
\includegraphics[width=1.99\columnwidth]{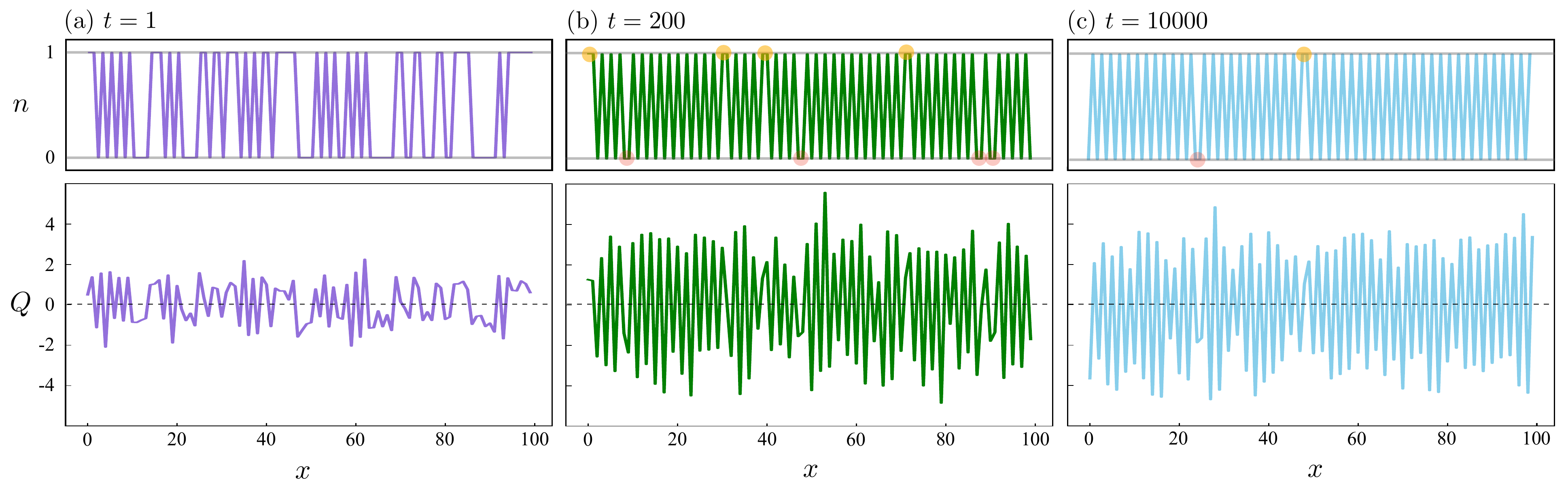}
\caption{Configuration snapshots of the quasi-classical Holstein model following a thermal quench, showing the electron occupation number  $n_i$ (top panels) and the corresponding local lattice distortion $Q_i$ (bottom panels) at three representative times: (a) $t = 1$, (b) $t=200$, and (c) $t=10000$.}
 \label{fig:configs}  \
\end{figure*}

The forces on the lattice variables remain given by Eqs.~(\ref{eq:force-total})--(\ref{eq:F_elastic}), with the quasi-equilibrium electron density now determined by Eq.~(\ref{eq:quasiclassical-electrondensity}). In this quasi-classical (q-cl) limit, the force at site-$i$ can be written explicitly in terms of the lattice distortion $Q_i$:
\begin{eqnarray}
	\label{eq:q-cl-F}
	F^{\scriptsize \mbox{q-cl}}_i = - k Q_i - \kappa \sum_{j \in \mathcal{N}(i)}  Q_j + \frac{g}{2} \tanh\left(\frac{g Q_i + \varepsilon_F}{2T}\right). \qquad
\end{eqnarray}
The dynamical evolution of the system in the adiabatic approximation is governed by the Langevin equation 
\begin{eqnarray}
	\label{eq:langevin-qcl}
	m\frac{d^2Q_i}{dt^2} = F^{\scriptsize \mbox{q-cl}}_i  - \gamma \frac{dQ_i}{dt} + \eta_i(t),
\end{eqnarray}
with the quasi-classical forces given in Eq.~(\ref{eq:q-cl-F}). This closed-form expression for the forces substantially reduces the computational cost of dynamical simulations compared with the full quantum dynamical approach described in Sec.~\ref{sec:semiclassical-Holstein}. The quasi-classical Holstein model is naturally characterized by an energy scale $\varepsilon = g Q_0 = k Q_0^2 / 2$, where $Q_0 = g / k$ denotes the characteristic amplitude of lattice distortion; see Appendix~\ref{apd:DimensionalAnalysis} for details of this dimensional analysis. Throughout the following, all energies and temperatures are expressed in units of $\varepsilon$. A convenient unit of time is set by the natural oscillation period $t_0 = 2\pi / \omega_0$, where $\omega_0 = \sqrt{k / m}$ is the fundamental frequency of the local harmonic oscillators. All simulation times presented below are measured in units of $t_0$. 

Using this formulation, we carried out extensive simulations of thermal quenches in the quasi-classical Holstein model. Specifically, the system was initialized in a random configuration of ${Q_i}$ and then suddenly coupled to a thermal reservoir at a low temperature $T = 0.032$.
Fig.~\ref{fig:configs} presents snapshots of the electron occupation $n_i$ and lattice displacement $Q_i$ at three representative times following the thermal quench, obtained from Langevin-dynamics simulations. Immediately after the quench, both fields remain strongly disordered, reflecting the high-temperature initial state. As the system evolves, short-wavelength fluctuations begin to relax and coarsen, and the lattice field develops locally correlated distortions. At late times, the dynamics approaches a quasi-steady low-temperature regime characterized by a well-formed charge-density-wave pattern in $n_i$ accompanied by large-amplitude, staggered distortions in $Q_i$.

During this coarsening process, the lattice displacement organizes into CDW domains whose amplitudes gradually approach their equilibrium values. The magnitude of the distortion is notably suppressed at the interfaces between domains---kinks where $Q_i Q_{i+1} < 0$. This suppression reflects the fact that the excess elastic and coupling energy during ordering becomes localized at these domain walls~\cite{Bray1994}. Under Langevin evolution, coarsening proceeds predominantly through kink diffusion and annihilation, leading to a monotonic reduction in the total number of kinks, and hence a decreasing kink density $\rho(t)$, as qualitatively illustrated in Fig.~\ref{fig:configs}.

To quantify the growth of the characteristic CDW domain size, we analyze the equal-time spatial correlation function of the lattice displacement field,
\begin{equation}\label{eqn:corrftn}
C_{ij}(t)=\langle Q_i(t) Q_j(t)\rangle-\langle Q_i(t)\rangle\langle Q_j(t)\rangle,
\end{equation}
where $i$ and $j$ label lattice sites. The averages $\langle \cdots \rangle$ are taken over all lattice sites and further averaged over an ensemble of independently quenched initial configurations to improve statistical accuracy.
A key advantage of working in the quasi-classical limit of the Holstein Hamiltonian is that the electronic occupations $n_i$ become strictly local functions of the lattice configuration. As a result, the evaluation of the electronic force scales as $\mathcal{O}(N)$, in stark contrast to the $\mathcal{O}(N^3)$ cost associated with real-space diagonalization of the electronic Hamiltonian. This reduction in computational complexity enables simulations of substantially larger system sizes, which is essential for probing CDW phase ordering: the coarsening process couples microscopic electronic energetics to emergent mesoscale dynamics, and a faithful characterization of the ordering regime requires access to large domains and long evolution times.
In this work, we simulate chains of length $N=10000$ and compute the correlation function using averages over 200 independent realizations. These choices ensure that both finite-size effects and statistical fluctuations are well controlled throughout the coarsening regime.

\begin{figure}
\includegraphics[width=0.9\columnwidth]{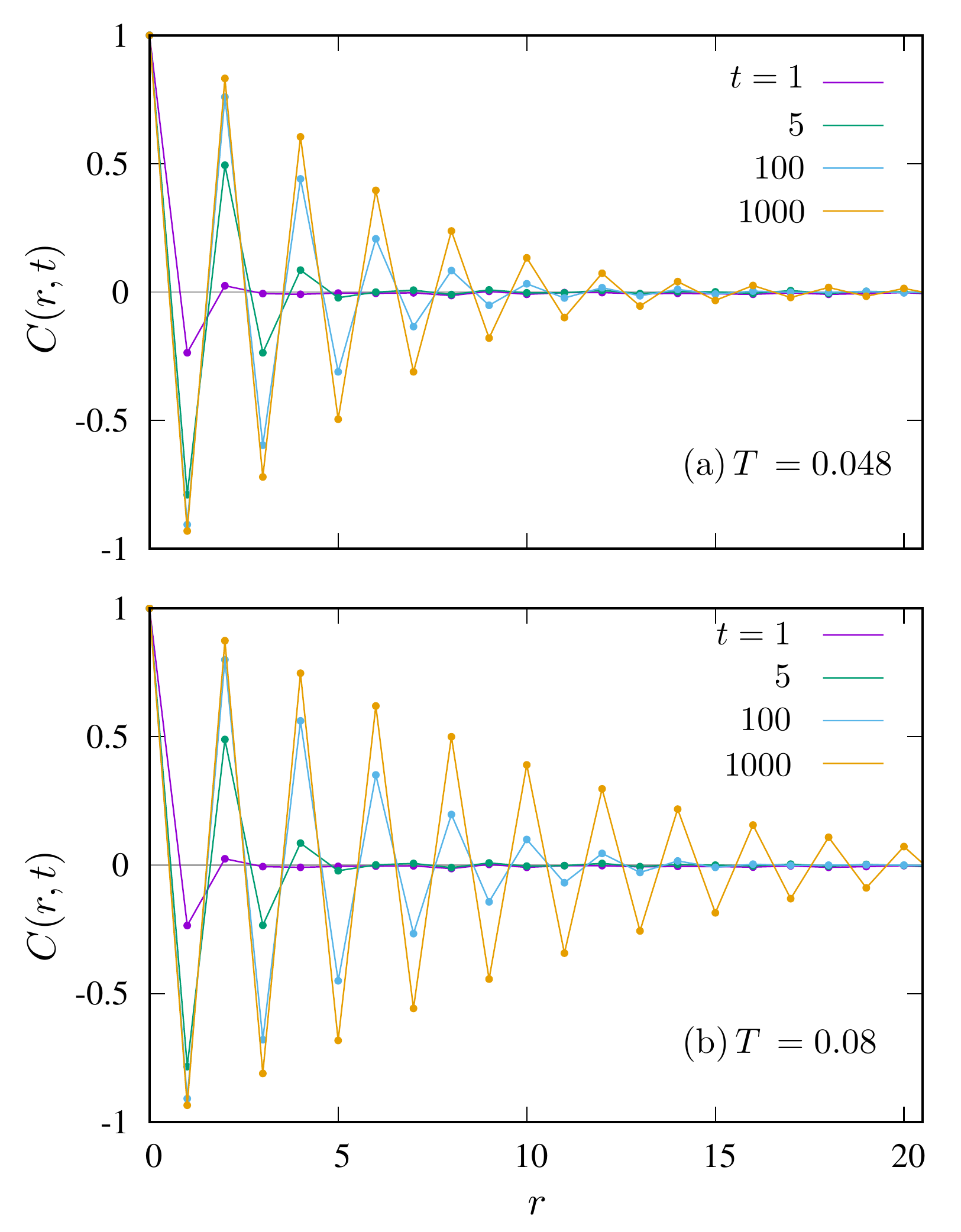}
\caption{Spatial correlation function of the lattice displacement field, $C(r,t)$, evaluated at several representative times following the thermal quench for two temperatures. Here $r = r_{ij}$ is the distance between a pair of sites-$i$ and $j$. Panels (a) and (b) show results for $T=0.048$ and $T=0.08$, respectively. }
\label{fig:corr}  \
\end{figure}

The spatial correlation functions in Fig.~\ref{fig:corr} provide a direct measure of CDW domain growth following the thermal quench. At early times ($t=1$ and $5$), the correlations decay almost immediately, indicating that the lattice remains strongly disordered. As time progresses, oscillatory correlations with alternating sign begin to emerge and extend over increasing distances, signaling the development of staggered lattice distortions and the onset of CDW domain formation. By $t=100$, these oscillations persist over several lattice spacings, reflecting the establishment of mesoscopic order.

At later times ($t=1000$), the oscillatory structure becomes more pronounced and longer-ranged, consistent with substantially larger domains and enhanced coherence. A comparison between the two temperatures highlights the role of thermal fluctuations: at $T=0.048$, the correlations are stronger and decay more slowly than at $T=0.08$, where thermal noise more effectively disrupts long-range coherence. Because true long-range Ising order is absent in one dimension at any finite temperature, the system ultimately remains disordered for all quenched temperatures, with only quasi-long-range CDW correlations developing at late times. Overall, the results capture the temperature-dependent coarsening dynamics of the quasi-classical Holstein chain and the gradual growth of CDW domains.
A convenient quantitative measure of this coarsening process is the time-dependent correlation length $\xi(t)$, which—as discussed in Sec.~\ref{sec:intro}—is inversely proportional to the kink density in one dimension, $\xi(t) \sim 1/\rho(t)$. To further clarify the growth of domains, we therefore examine the temporal evolution of the kink number and kink density.

\begin{figure}
\includegraphics[width=0.9\columnwidth]{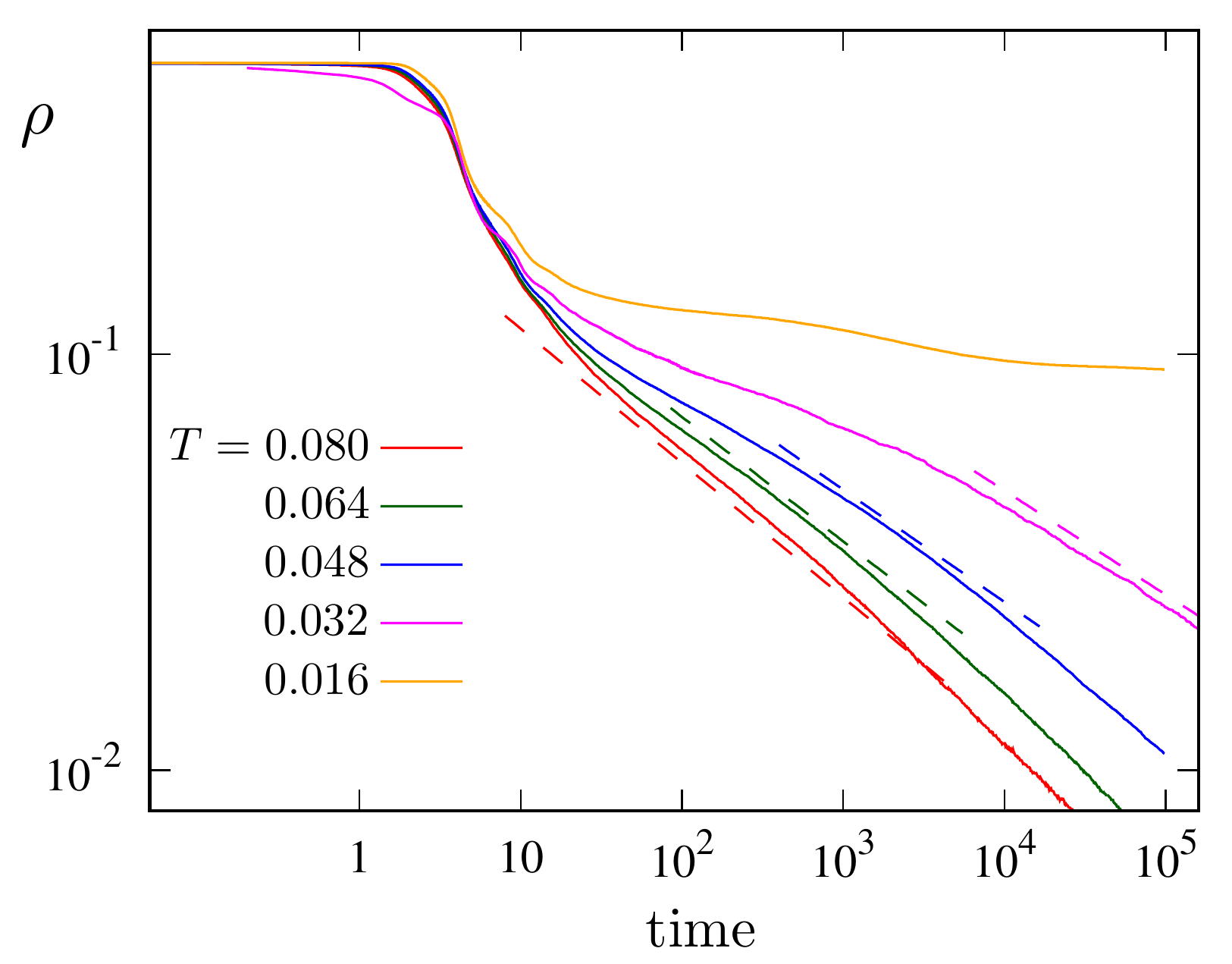}
\caption{Density of kinks $\rho$ versus time of the 1D quasi-classical Holstein model at varying quench temperatures. The dashed lines correspond to power-law decay $1/t^{1/(2+a)}$, with the nonlinear parameter $a = 1.1$, 1.35, 1.7, 2 for temperatures $T = 0.08$, 0.064, 0.048, 0.032 (from bottom to top). Simulations are performed on a chain of length $N = 10000$.}
\label{fig:nk-t-vs-T}  \
\end{figure}

Fig.~\ref{fig:nk-t-vs-T} shows the time dependence of $\rho(t)$ following a thermal quench to several temperatures. At the lowest temperature studied, $T = 0.016$, the kink density begins to plateau at late times, suggesting an eventual freezing of the coarsening process. For higher temperatures, however, the system after a brief initial transient enters a well-defined coarsening regime in which the kink density exhibits a clear power-law decay $\rho \sim t^{-1/z}$ for all temperatures studied. Importantly, the decay exponent is temperature dependent: lower-temperature quenches display systematically slower decay rates. Moreover, in every case the extracted exponents differ markedly from the canonical value $z = 2$ expected for diffusion-limited annihilation in one dimension. As discussed in detail below, this deviation indicates that kink motion is not governed by simple free diffusion but is instead strongly influenced by nonlinear, cooperative hopping dynamics intrinsic to the quasi-classical Holstein model.

\section{Cooperative Kink-Hopping Mechanism}

Here we first briefly review the phenomenological diffusion--reaction framework for describing kink dynamics during one-dimensional coarsening. Following a quench into a symmetry-broken phase, the system consists of alternating domains separated by kinks (domain walls), which behave as deconfined topological defects. At late times, the motion of kinks is well approximated by unbiased diffusion with diffusion constant $D$. Whenever two neighboring kinks collide, they annihilate via the reaction $A + A \rightarrow \varnothing$, corresponding to the disappearance of a domain of vanishing size~\cite{Sheu1990, Kang1985, Habib1999,hoogeman2000, kawasaki1982,kawasaki1983,bramson1980,Richard1981}. The essential assumption of \emph{diffusion-limited annihilation} is that the microscopic annihilation process is instantaneous compared to the diffusive time required for two kinks to encounter each other. As a result, the reaction kinetics is governed entirely by the first-passage properties of diffusing random walkers.

A coarse-grained description of this process is provided by the reaction--diffusion equation for the local kink density $\rho(x,t)$~\cite{heidari2023,Smoller1994, KPP1937, Boon2012},
\begin{equation}
    \label{eq:RD-kinks}
    \frac{\partial \rho}{\partial t}
    = D \frac{\partial^2 \rho}{\partial x^2}
      - K \rho^2,
\end{equation}
where $K$ is the bare reaction rate. The quadratic term reflects the fact that two kinks must coincide spatially in order to annihilate. A naive mean-field treatment assumes spatial uniformity, producing the ordinary differential equation $d\rho/dt = -K\rho^2$ with solution $\rho(t) \sim 1/t$. This mean-field kinetics is valid only when diffusion efficiently mixes particles and eradicates spatial anticorrelations--a condition that requires spatial dimensions $d>2$, where random walks are \emph{transient} and particles do not revisit the same region indefinitely.

In one dimension, however, diffusion is \emph{recurrent}: two random walkers meet with probability one, and strong spatial anticorrelations emerge between surviving kinks. Under these conditions, the mean-field factorization fails, and Eq.~\eqref{eq:RD-kinks} no longer captures the true asymptotic kinetics. Instead, the dynamics becomes controlled by diffusive first-passage events, leading to a growing inter-kink spacing $\xi(t) \sim \sqrt{D t}$, and hence a kink density
\begin{equation}
    \label{eq:power-law-0.5}
    \rho \sim t^{-1/2}.
\end{equation}
This diffusion-limited behavior may also be obtained from a modified mean-field argument. In $d$ spatial dimensions, the Smoluchowski theory of diffusion to an absorbing sphere predicts a diffusion-limited reaction rate $K \sim D R^{d-2}$, where $R$ is the effective capture radius of the absorber \cite{Krapivsky_book,Ovchinnikov1989,Smoluchowski1917,Chandrasekhar43}. In dimensions $d>2$, this yields a constant reaction coefficient, recovering the mean-field law $\rho \sim 1/t$. In contrast, in 1D diffusion is recurrent, so encounters are entirely controlled by first-passage events; effectively, the ``reaction radius'' is set by the typical inter-kink spacing, $R \sim \xi \sim 1/\rho$. Substituting this into the Smoluchowski scaling gives an effective reaction rate $K_{\mathrm{eff}}  \sim D\rho$, and hence an effective rate equation
\begin{equation}
    \label{eq:effective-rate-eq}
    \frac{d\rho}{dt} \sim - D \rho^3,
\end{equation}
whose solution again yields a power-law decay consistent with Eq.~\eqref{eq:power-law-0.5}. Thus, the characteristic $t^{1/2}$ coarsening law in one-dimensional Ising-type systems follows directly from the
diffusion-limited annihilation of kinks.

\begin{figure}
\includegraphics[width=0.95\columnwidth]{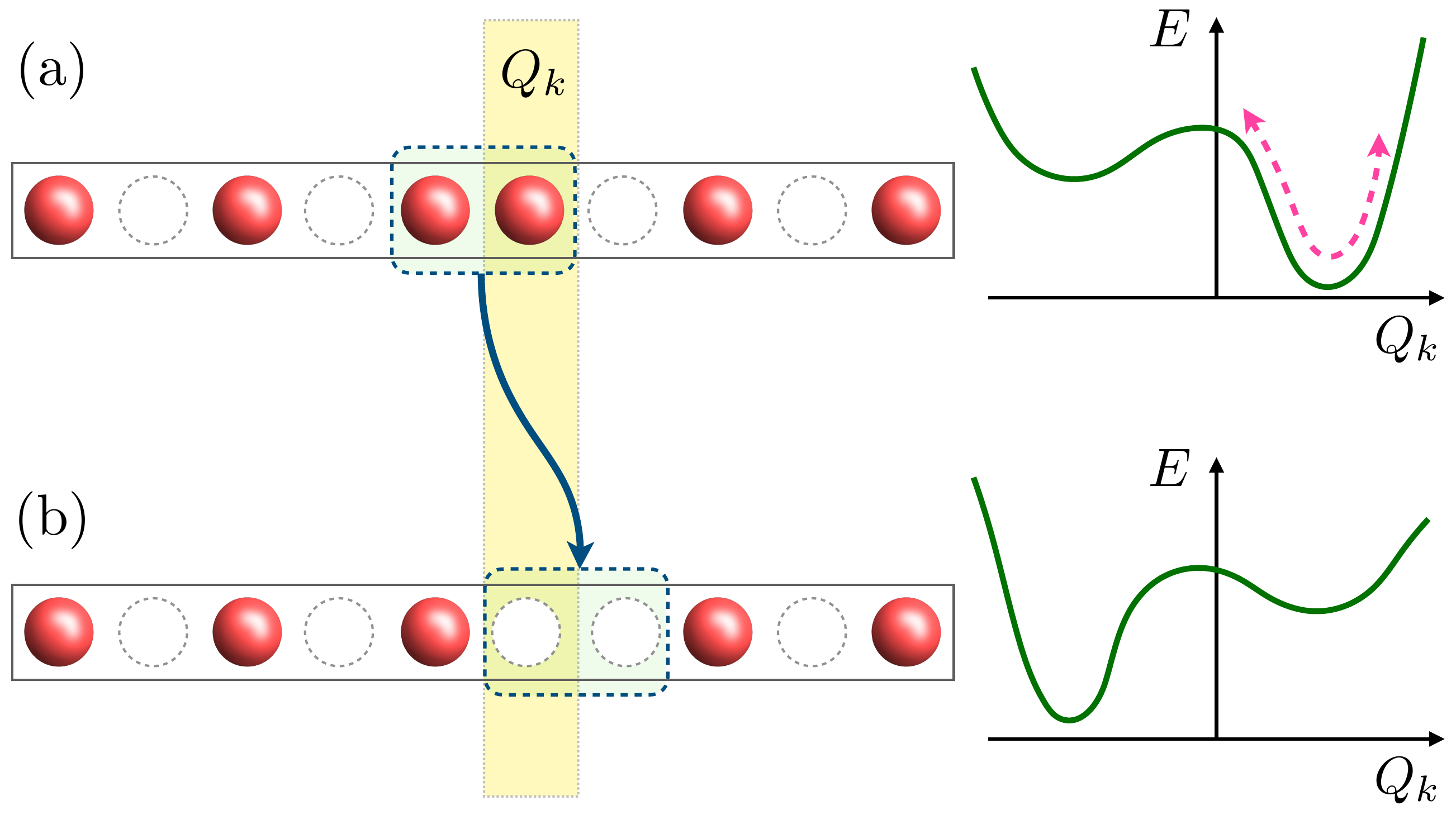}
\caption{Schematic illustration of the thermally activated hopping of a single kink. (a) Thermal fluctuations drive the local lattice displacement $Q_k$ away from the positive-$Q$ minimum associated with an occupied site, enabling the system to overcome the energy barrier separating the two minima (occupied versus empty). (b) After crossing the barrier, $Q_k$ relaxes into the negative-$Q$ minimum corresponding to an empty site, resulting in the kink shifting by one lattice constant to the right and completing a single hopping event.}
\label{fig:hopping-schematic}  
\end{figure}

Next we turn to the microscopic hopping dynamics of kinks in the quasi-classical Holstein model. To build intuition, we first consider the simplified limit in which the electron occupation is strictly binary, $n_i \in \{0,1\}$, as is appropriate at very low temperatures. A kink located on the bond $(i,i+1)$—which interpolates between the two CDW ground states related by the underlying $Z_2$ symmetry—corresponds to a configuration in which the two adjacent sites are either both occupied, $n_i = n_{i+1} = 1$, or both empty, $n_i = n_{i+1} = 0$. For a kink of the former type, a naive picture of kink motion arises from shifting one of the electrons by a single lattice spacing to the right:
\begin{eqnarray*}
   & & \cdots \,\,  \occsite\ \empsite\ \occsite\ \empsite\ \occsite\ \empsite\ \occsiteRed\ \occsiteRed\ \empsite\ \occsite\ \empsite\ \occsite\ \empsite\ \occsite\ \empsite\ \,\, \cdots  \\
   & & \hspace{68pt} \Downarrow \\
   & & \cdots \,\,  \occsite\ \empsite\ \occsite\ \empsite\ \occsite\ \empsite\ \occsite\ \empsite\ \occsiteRed\ \occsiteRed\ \empsite\ \occsite\ \empsite\ \occsite\ \empsite\ \,\, \cdots
\end{eqnarray*}
In this process, the electron hop effectively moves the kink by two lattice constants to the right. Interestingly, detailed examination of our simulation snapshots shows that kinks actually move only one lattice constant through the following process:
\begin{eqnarray*}
   & & \cdots \,\,  \occsite\ \empsite\ \occsite\ \empsite\ \occsite\ \empsite\ \occsiteRed\ \occsiteRed\ \empsite\ \occsite\ \empsite\ \occsite\ \empsite\ \occsite\ \empsite\ \,\, \cdots  \\
   & & \hspace{68pt} \Downarrow \\
   & & \cdots \,\,  \occsite\ \empsite\ \occsite\ \empsite\ \occsite\ \empsite\ \occsite\ \empsiteRed\ \empsiteRed\ \occsite\ \empsite\ \occsite\ \empsite\ \occsite\ \empsite\ \,\, \cdots
\end{eqnarray*}
Here the kink shifts to the right by one site, while simultaneously transforming from a $\,\occsiteRed\ \occsiteRed\,$ type to a $\,\empsiteRed\ \empsiteRed\,$ type. For convenience, we shall refer to this as a type-A hopping, while the reverse transformation $\,\empsiteRed\ \empsiteRed\,\to \,\occsiteRed\ \occsiteRed\,$ is called type-B hopping. 
It is worth noting that this one-step kink hop is fundamentally a single-site process, localized at the position where the electronic occupation flips from $n = 1$ to $0$. A schematic depiction of the underlying microscopic mechanism is shown in Fig.~\ref{fig:hopping-schematic}, whose right panels illustrate the corresponding evolution of the effective potential energy landscape for the local displacement $Q_k$ before and after the hop.

Before the hopping event, the local lattice displacement $Q_k$ resides in the deeper, positive-$Q$ minimum of its double-well potential, corresponding to an occupied electronic site. This minimum is separated by an energy barrier $\Delta$ from the shallower negative-$Q$ minimum associated with an empty site. At finite temperatures, thermal fluctuations intermittently push $Q_k$ away from the positive-$Q$ minimum, allowing the system to surmount the barrier and reach the transition point between the two wells. Once the barrier is crossed, $Q_k$ rapidly relaxes into the negative-$Q$ minimum, thereby converting the local electronic occupation from $n=1$ to $n=0$ and shifting the kink by one lattice constant to the right. This thermally activated process thus represents the elementary step of kink propagation. The corresponding hopping rate follows an Arrhenius form~\cite{Krapivsky_book}, leading to a diffusion constant that scales as $D \sim e^{-\Delta / T}$.

Importantly, the thermally activated nature of the local $Q_k$ transition also clarifies why the two-step kink-hopping scenario is statistically far less probable. In this alternative process, the kink would move by two lattice constants only if two adjacent sites undergo coordinated barrier crossings: one site must transition from the positive-$Q$ minimum to the negative-$Q$ minimum, while its neighbor must simultaneously undergo the opposite transition from negative to positive $Q$. Each of these transitions individually requires overcoming an energy barrier of height $\Delta$, and the probability of both occurring in concert is suppressed by an additional Arrhenius factor. In effect, the two-step process demands a rare, correlated fluctuation involving two independent thermally activated events at two {\em adjacent} sites. As a result, the cooperative two-site transition is exponentially less probable, and kink propagation is therefore dominated by the single-step process identified in our simulations.

\begin{figure}
\includegraphics[width=0.95\columnwidth]{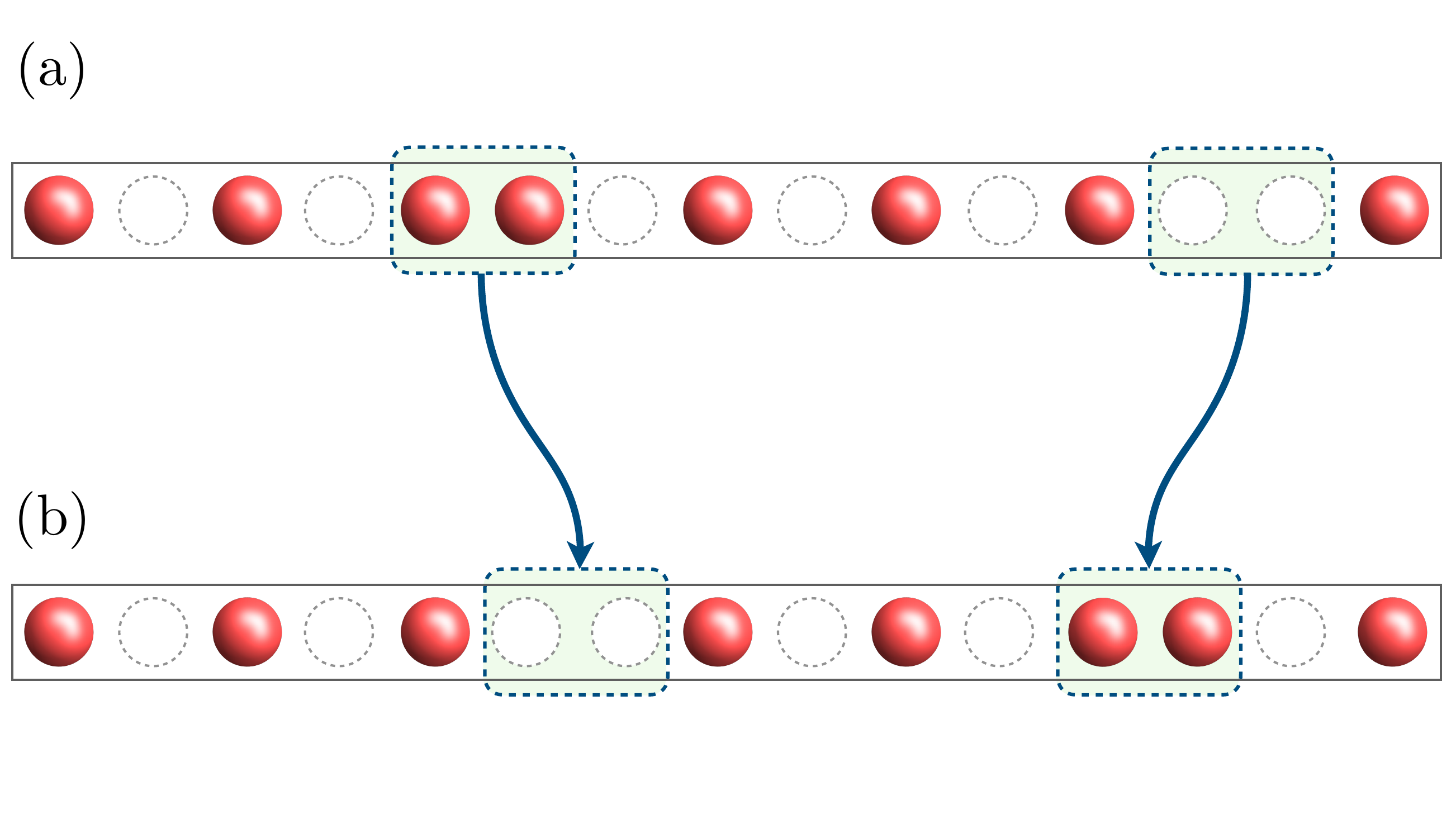}
\caption{Schematic illustration of the cooperative hopping of two kinks required to conserve the total electron number. As shown by comparing panels (a) and (b), a coordinated two-kink event---consisting of a type-A hop for the left kink and a type-B hop for the right kink---simultaneously shifts both kinks while converting each into its opposite type, thereby preserving the net electron number of the system.}
\label{fig:cooperative-hopping}  \
\end{figure}

On the other hand, a subtle complication with the single-kink hopping scenario shown in Fig.~\ref{fig:hopping-schematic} is that it does not conserve the total electron number. In the illustrated type-A process, the transformation of a $\,\occsiteRed\ \occsiteRed\,$ kink into a $\,\empsiteRed\ \empsiteRed\,$ kink effectively removes one electron from the system, while the reverse type-B process would correspondingly add an electron. For systems that do not exchange particles with an external reservoir---in particular, those described within the canonical ensemble---the total electron number must remain fixed (here, at half-filling). This constraint implies that physically allowed kink motion must proceed through cooperative processes involving both kink types. As shown in Fig.~\ref{fig:cooperative-hopping}, a coordinated two-kink event---comprising a type-A hop for the left kink and a type-B hop for the right kink---advances both kinks simultaneously while converting each into its opposite type. This joint motion preserves the net electron number exactly, ensuring that kink propagation takes the form of correlated rather than independent hopping. Importantly, such cooperative hopping behavior is directly observed and confirmed in our numerical simulations.

Even for systems described by the grand-canonical ensemble, where electron exchange with a reservoir is formally permitted, the cooperative mechanism remains essential. In our grand-canonical Langevin simulations, the on-site electron density Eq.~(\ref{eq:quasiclassical-electrondensity}) is evaluated at a fixed chemical potential $\varepsilon_F$ chosen to enforce approximate half-filling. As discussed in Appendix~\ref{sec:grand-canonical}, the resulting kink density $\rho(t)$ exhibits power-law coarsening analogous to Fig.~\ref{fig:nk-t-vs-T}, with temperature-dependent exponents $z$ that deviate strongly from the diffusion-limited value $z = 2$. This behavior arises because, although the particle number is not strictly fixed in the grand-canonical setting, maintaining the chemical potential close to half-filling still requires kink motion to proceed through cooperative hopping processes rather than isolated single-kink events. 

The cooperative hopping mechanism has an important kinetic consequence: because each correlated hop involves one kink of each type, the overall hopping rate depends on the number of such available pairs. Let $\mathcal{N}_k(t)$ denote the total number of kinks at time $t$. Statistically, the two kink species occur with roughly equal frequency, so each appears in number $\mathcal{N}_k/2$. The number of possible cooperative pairs therefore scales as $(\mathcal{N}_k / 2)^2$, or equivalently as $\rho^2$ in terms of kink density. This combinatorial effect may be incorporated phenomenologically by promoting the effective diffusion coefficient to a density-dependent form, $D \sim D_0 \rho^2$ in addition to the Arrhenius factor $D_0(T) \sim e^{-2\Delta/T}$ associated with the underlying barrier-crossing events. 

It is worth noting that the above consideration also explains why the two-step kink hopping described above--- where one electron at the kink-site hops to its immediate neighbor---is statistically highly improbable. Such a process corresponds to the limiting case in which a type-A and a type-B hop occur adjacent to one another at precisely the right instant. 

At finite temperatures, however, lattice sites acquire partial occupancies according to Eq.~(\ref{eq:quasiclassical-electrondensity}), especially for sites whose energies lie near the Fermi level. Such partial filling introduces additional pathways for correlated motion and can modify the combinatorial structure of the cooperative hopping processes. To accommodate these more general scenarios, we therefore allow a generic power-law dependence of the effective diffusivity on the kink density,
\begin{eqnarray}
	D = D_0 \rho^a,
\end{eqnarray}
with an exponent $a$ not fixed a priori. Nonlinear density-dependent diffusion of this form has been extensively studied in diverse physical systems~\cite{machida1978,stewart1982,white1984,Genghmun1989,kuntz2003, Krapivsky12, hernandez2008,Lutsko2008,Cornell1991,Majumdar1995,Sphon1991}. In the present context, however, we identify a microscopic origin unique to our system: the nonlinearity arises directly from the fermionic statistics encoded in the Fermi-Dirac occupations governing the cooperative kink hopping processes.


This nonlinear diffusivity also implies a different asymptotic coarsening dynamics. The additional power-law dependence of the effective diffusion coefficient modifies the effective rate equation Eq.~(\ref{eq:effective-rate-eq}) to $d\rho/ dt \sim -D_0 \rho^{a + 3}$, which is readily integrated to yield a power-law decay 
\begin{equation}\label{eqn:kink-decay-exponent}
    \rho \sim t^{-1/(2+a)}.
\end{equation}
corresponding to a dynamical exponent $z = 2 + a$. Thus, departures from the diffusion-limited value $z = 2$ directly reflect the strength of the density-dependent cooperative hopping encoded in the exponent~$a$.

Our Langevin simulation results in Fig.~\ref{fig:nk-t-vs-T} provide clear evidence for this mechanism. The measured coarsening exponents exhibit a pronounced temperature dependence, with $z(T)$ increasing systematically as the temperature decreases. This trend indicates that kink motion becomes progressively more constrained at low $T$, where cooperative hopping dominates over independent single-kink motion. Notably, at $T = 0.032$---the lowest temperature at which the decay of the kink density can still be well described by a power law---the best fit corresponds to an effective exponent $a = 2$, precisely the value expected in the binary-occupation limit where only strictly correlated kink pairs can participate in hopping.

The monotonic increase of $a(T)$ with decreasing temperature reflects the growing importance of cooperative barrier-crossing events as thermal fluctuations diminish and the electron occupation approaches the binary limit. In this regime, kink motion becomes progressively more dependent on the simultaneous availability of compatible neighboring configurations, enhancing the effective nonlinearity of the diffusion process and thus producing the observed temperature-dependent coarsening exponents.

\begin{figure}
\includegraphics[width=0.99\columnwidth]{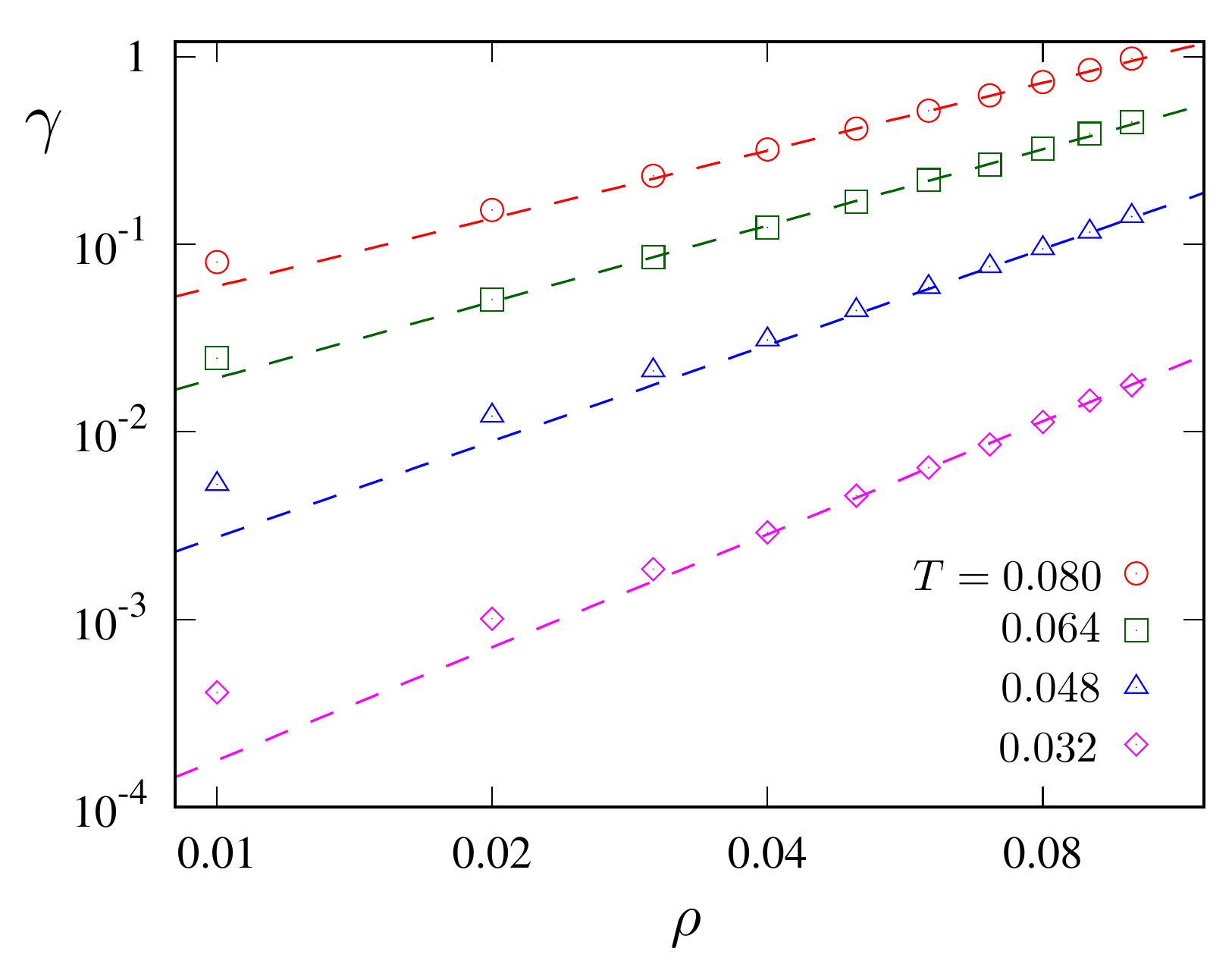}
\caption{Hopping rate $\gamma$ as a function of kink density $\rho$ obtained from Langevin dynamics simulations at various temperatures. The dashed lines are fittings with power-law $\gamma \sim \rho^a$, where the exponent $a = 1.1$, 1.35, 1.7 and 2 for temperatures $T = 0.08$, 0.064, 0.048, 0.032 (from top to bottom). Simulations are performed on a chain of length $N = 1000$.}
\label{fig:hopping-rate}  
\end{figure}

To more directly validate the proposed density-dependent nonlinear diffusion mechanism, we performed Langevin dynamics on specially prepared configurations designed to isolate correlated hopping processes. Specifically, we initialize the electron occupations ${ n^{(0)}_i }$ as an evenly spaced array of kinks separated by $\ell$ lattice constants, effectively forming a one-dimensional “kink lattice.” The corresponding initial lattice configuration ${ Q_i }$ is obtained by running a constrained Langevin dynamics in which the electron occupations are held fixed in this prescribed array pattern (i.e., the ${ n^{(0)}_i }$ are not updated). After the lattice variables have reached thermal equilibrium under this constraint, we release the system and perform standard, fully coupled Langevin dynamics, in which the electronic occupations evolve self-consistently with the instantaneous lattice configuration through Eq.~(\ref{eq:quasiclassical-electrondensity}).

We then monitor the dynamics by counting individual kink-hopping events, which are identified by sign changes of $Q$ at one of the sites comprising a kink. The simulations are run either for a fixed amount of time or until the first kink-annihilation event occurs (which reduces the total kink number by two). From these trajectories, we define a hopping rate $\gamma$ as the total number of observed kink-hopping events divided by the simulation time and the number $\mathcal{N}_k = N/\ell$ of kinks present. By repeating such simulations for different spacings $\ell$ of the initial kink lattice, we obtain the density-dependent hopping rate as a function of the kink density $\rho = 1/ \ell$ at various temperatures. Notably, the effective diffusion coefficient $D_{\rm eff}$ is expected to be proportional to this hopping rate, allowing us to extract $D_{\rm eff}$ directly from the Langevin data.

Fig.~\ref{fig:hopping-rate} shows the kink-hopping rate $\gamma$ as a function of kink density for several temperatures. In analyzing the density dependence, we focus on the higher-density regime where finite-size effects—significant at low kink density—are negligible. At each temperature, the hopping rate increases systematically with density, consistent with a power-law-like behavior from which an effective exponent $a(T)$ can be extracted. A clear trend emerges: $a(T)$ decreases with increasing temperature. From low temperature ($T = 0.016$) to high temperature ($T = 0.08$), the exponent approaches unity ($a \approx 1.1$), indicating that the hopping rate becomes only weakly dependent on kink density. This reduction in the exponent reflects the enhanced role of thermal fluctuations at higher temperatures: thermal noise more readily produces local configurations that satisfy the electron-number constraint, enabling kink motion through cooperative events involving only a few neighboring kinks rather than requiring a high overall kink density. As a result, the mobility of an individual kink becomes less strongly controlled by the number of other kinks present in the system. The exponent $a$ extracted from the hopping-rate analysis also agree very well with those obtained from the power-law decay shown in Fig.~\ref{fig:nk-t-vs-T}.

\section{Summary and Outlook}

\label{sec:summary}

In summary, we have conducted a comprehensive numerical investigation of charge density wave (CDW) coarsening in the one-dimensional, half-filled Holstein model. Following a thermal quench, Ising-like domains form and grow through the dissipative dynamics of the lattice, with kinks acting as the topological defects mediating domain-wall motion. Across a wide range of model parameters, we find regimes in which kink motion becomes strongly hindered, producing significantly slower coarsening than the standard diffusion-controlled behavior of one-dimensional Ising systems. To uncover the microscopic origin of this anomalous kinetics, we turned to the quasi-classical limit of the Holstein model, which preserves the essential fermionic constraints while rendering the electronic occupations strictly local and analytically transparent.

Large-scale simulations show that the slow coarsening persists in this simplified limit, providing a clean setting in which to identify the fundamental mechanism: under electron-number conservation, a single kink hop necessarily changes the local charge, and therefore cannot occur in isolation. Instead, kink motion must proceed through a cooperative two-kink process in which a hop of one kink is paired with a compensating hop of another of the opposite type. This requirement, which becomes explicit only in the quasi-classical formulation, suppresses independent defect motion and leads to a strong dependence of kink mobility on kink density and temperature. Direct numerical measurements of kink hopping rates confirm this cooperative character and establish a clear correspondence between microscopic electron-number constraints and the emergent slow phase-ordering dynamics.

The cooperative-hopping mechanism provides a powerful conceptual framework for understanding defect dynamics not only in the quasi-classical model, but also in the full semi-classical Holstein system. Although finite electron hopping introduces additional channels for defect motion and partial screening, the essential constraint---that defect motion locally redistributes electronic charge---still influences kink dynamics and can inhibit independent domain-wall motion. Similar considerations apply even more broadly to electron-lattice systems in higher dimensions: wherever topological defects couple to local electronic occupations, defect transport may require coordinated electronic rearrangements, resulting in constrained mobility, slow coarsening, or glassy relaxation. The quasi-classical Holstein chain thus serves as a uniquely transparent model for identifying these fundamental mechanisms, and it provides a foundation for future studies of constraint-driven dynamics in higher-dimensional CDW materials, and other strongly coupled electron-lattice models.

\begin{acknowledgments}
The work was supported by the US Department of Energy Basic Energy Sciences under Contract No. DE-SC0020330.  The authors also acknowledge the support of Research Computing at the University of Virginia.
\end{acknowledgments}

\appendix

\section{Dimensional Analysis}

\label{apd:DimensionalAnalysis}

Here, we perform the dimensional analysis of the quasi-classical Holstein model and derive the corresponding equation of motion for the lattice. To nondimensionalize the model, we first identify the natural scales of the lattice sector. The on-site elastic term $(k/2)Q_i^2$ defines a local harmonic-oscillator frequency $\omega_0$ and the corresponding characteristic time scale:
\begin{equation}
\omega_0 = \sqrt{k/m}, \qquad t_0 = 2\pi / \omega_0.
\end{equation}
We therefore introduce the dimensionless time variable 
\begin{eqnarray}
	\tilde t = \omega_0 t.
\end{eqnarray}
The characteristic amplitude of lattice distortion follows from minimizing the competing electron-lattice coupling and elastic energies, $-g n_i Q_i$ and $(k/2)Q_i^2$, respectively. Assuming $n_i \sim \mathcal{O}(1)$, the resulting displacement scale is
\begin{equation}
Q_0 = g/k .
\end{equation}
The associated momentum scale $P_0 = m\omega_0 Q_0$ then motivates the dimensionless variables
\begin{equation}
\tilde Q_i = Q_i/Q_0,
\qquad
\tilde P_i = P_i/P_0 .
\end{equation}
The characteristic distortion $Q_0$ also allows us to introduce a typical energy scale associated with electron-lattice coupling:
\begin{eqnarray}
	\varepsilon = g Q_0 = g^2/k
\end{eqnarray}
The electron Hamiltonian, for example, can be expressed as $\hat{\mathcal{H}}_e = \varepsilon \sum_i (\hat{n}_i - 1/2) \tilde{Q}_i$. In terms of these dimensionless variables, the Langevin equation becomes
\begin{eqnarray}
\label{eq:eom}
	\frac{d^2 \tilde Q_i}{d\tilde t^2} = \left(n_i - \frac{1}{2} \right) - \tilde Q_i - \tilde\kappa \sum_{j\in\mathcal{N}(i)} \tilde Q_j
	- \tilde\gamma\frac{d\tilde Q_i}{d\tilde t}	+ \tilde\eta_i(\tilde t), \nonumber \\
\end{eqnarray}
where $\tilde\kappa=\kappa/k$ and $\tilde\gamma=\gamma/(m\omega_0)$ denote the dimensionless elastic coupling and damping coefficient. The Gaussian white-noise term satisfies
\begin{eqnarray}
\langle \tilde{\eta}_i(t) \rangle 
&=& 0, \nonumber\\[4pt]
\langle \tilde{\eta}_{i a}(t)\, \tilde{\eta}_{j b}(t') \rangle 
&=& \tilde{D}\,
\delta_{ij}\, \delta_{ab}\, \delta(\tilde t - \tilde t'), 
\end{eqnarray}
with a dimensionless diffusion constant $\tilde D = 2 \tilde{\gamma} \tilde{T}$, where the dimensionless temperature $\tilde T = T / \varepsilon$ is measured in units of the typical energy scale introduced above. In the simulations presented here we use $\tilde\gamma=0.16$, and $\tilde\kappa=0.3$ in Eq.~(\ref{eq:eom}), while the dimensionless noises $\tilde{\eta}_i$ are determined by normalized temperature $\tilde T$. The equations of motion are integrated using a velocity-Verlet scheme with time step $dt = 0.01$.

\section{quench dynamics with a constant chemical potential}

\label{sec:grand-canonical}

To test the robustness of the microscopic cooperative-hopping mechanism and its impact on coarsening, we also examined the relaxation dynamics in a grand-canonical setup, where the total electron number is allowed to fluctuate while the chemical potential is held fixed. All microscopic parameters of the Hamiltonian are kept identical to the canonical simulations; the only change lies in replacing the half-filling constraint with a constant Fermi level $\varepsilon$. This ensemble allows electrons to enter or leave the system through the reservoir, thereby relaxing the global particle-number constraint while retaining the same local electron-lattice energetics.

\begin{figure}
\includegraphics[width=0.99\columnwidth]{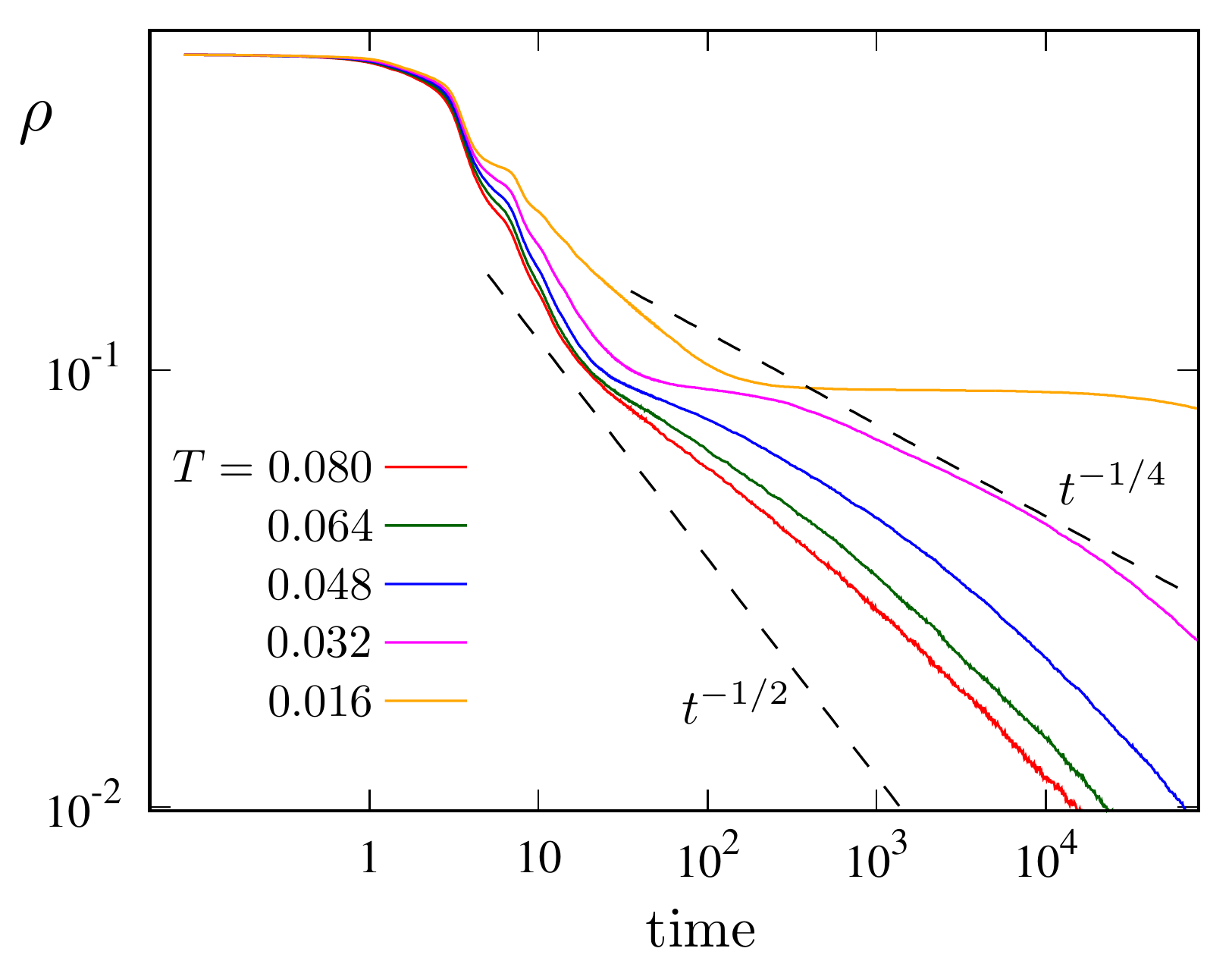}
\caption{Density of kinks $\rho$ versus time of the 1D quasi-classical Holstein model with a fixed chemical potential. Simulations are performed on a chain of length $N = 10000$.}
\label{fig:nk-t-vs-T-fixed_mu}  \
\end{figure}

Fig.~\ref{fig:nk-t-vs-T-fixed_mu} shows the kink-density decay $\rho(t)$ over a range of temperatures in the grand-canonical ensemble. As in the canonical case, the kink density exhibits a clear power-law decrease, $\rho \sim t^{-1/z}$, with a dynamical exponent $z(T)$ that varies systematically with temperature. At high temperature ($T = 0.08$), the extracted exponent exceeds the diffusion-limited value $z = 2$, confirming that the coarsening remains slower than what would be expected from independent random-walk dynamics. At lower temperatures, such as $T = 0.032$, the exponent remains below $z = 4$ but still significantly above $z = 2$, closely paralleling the trends observed in the canonical ensemble. These results demonstrate that the anomalously slow coarsening behavior persists even when global electron-number conservation is relaxed.

This grand-canonical analysis sheds light on the role of the chemical potential in controlling the kink dynamics. Fixing $\varepsilon_F$ enforces an average filling close to one-half but does not eliminate the local constraint inherent in kink motion. A single kink hop necessarily changes the occupation of a specific site, and such a local event must still satisfy electron-number conservation on the timescale of the hop itself. In practice, this means that even in the presence of a reservoir, kink motion continues to rely on cooperative two-kink processes that preserve the instantaneous local charge balance. Consequently, the same microscopic mechanism that slows coarsening in the canonical ensemble remains operative here. The persistence of temperature-dependent coarsening exponents in the grand-canonical simulations thus provides strong additional support for the cooperative-hopping picture and confirms that it is a robust and intrinsic feature of the quasi-classical Holstein dynamics.

\bibliography{ref}
\end{document}